\documentstyle[twoside,fleqn,espcrc2]{article}
\input{epsf}

\newcommand{\AmS}{{\protect\the\textfont2
  A\kern-.1667em\lower.5ex\hbox{M}\kern-.125emS}}

\def\g{$\gamma$}

\def\asca{{\it ASCA}}

\def\xte{{\it RXTE}}

\def\lh{$l_{\rm h}$}
\def\lth{$l_{\rm th}$}
\def\lnth{$l_{\rm nth}$}
\def\ls{$l_{\rm s}$}
\def\taup{$\tau_{\rm p}$}
\def\tauT{$\tau_{\rm T}$}
\def\kTth{$kT_{\rm th}$}
\def\Ts{$T_{\rm s}$}
\def\kTs{$kT_{\rm s}$}
\def\O2p{$\Omega/2\pi$}
\def\Gin{$\Gamma_{\rm inj}$}
\def\Rin{$R_{\rm in}$}
\def\Rg{$R_{\rm g}$}
\def\kTin{$kT_{\rm in}$}
\def\Ledd{$L_{\rm Edd}$}
\def\chisq{$\chi^2$}

\newbox\grsign \setbox\grsign=\hbox{$>$} \newdimen\grdimen %
\grdimen=\ht\grsign
\newbox\simlessbox \newbox\simgreatbox \newbox\simpropbox
\setbox\simgreatbox=\hbox{\raise.5ex\hbox{$>$}\llap
     {\lower.5ex\hbox{$\sim$}}}\ht1=\grdimen\dp1=0pt
\setbox\simlessbox=\hbox{\raise.5ex\hbox{$<$}\llap
     {\lower.5ex\hbox{$\sim$}}}\ht2=\grdimen\dp2=0pt
\setbox\simpropbox=\hbox{\raise.5ex\hbox{$\propto$}\llap
     {\lower.5ex\hbox{$\sim$}}}\ht2=\grdimen\dp2=0pt
\def\simgreat{\mathrel{\copy\simgreatbox}}
\def\simless{\mathrel{\copy\simlessbox}}

\title{Thermal/non-thermal model of Cyg X-1 in the soft state}

\author{M. Gierli\'nski\address{Jagiellonian University, Astronomical 
Observatory, Orla 171, 30-244 Krak\'ow, Poland},
A. A. Zdziarski\address{Copernicus Astronomical Center, Bartycka 18, 00-716 Warszawa, Poland},
P. S. Coppi\address{Astronomy Department, Yale University, P.O. Box 208101, New Haven, CT 062520-8101, USA},
J. Poutanen\address{Uppsala Observatory, Box 515, 75120 Uppsala, Sweden},
K. Ebisawa\address{NASA/Goddard Space Flight Center, Code 660.2, Greenbelt, MD 20771, USA}
and W. N. Johnson\address{E. O. Hulburt Center for Space Research, Naval Research Laboratory, Washington, DC 20375, USA}
}

\begin{document}

\begin{abstract}

\end{abstract}

\maketitle

\section{INTRODUCTION}

Cyg X-1 is the best-studied Black Hole Candidate. Its X-ray/$\gamma$-ray spectra undergo bimodal transitions between the two states: the hard one (so-called `low'), in which the X-ray spectrum is hard ($\alpha \sim$ 0.6) and the soft one (`high'), dominated by a soft X-ray component with a steeper ($\alpha \sim 1.5$) hard tail. Most of the time the source spends in the hard state.

We analyze two groups of 1996 observations of Cyg X-1. The first one consists of a simultaneous \asca/\xte\/ observation on May 30. The second one contains six simultaneous \xte/OSSE observations on June 17--18.

The \xte\/ data come from the public archive. A 2\% systematic error has been added to each PCA channel to represent calibration uncertainties. Since dead-time effects of the HEXTE clusters are not yet fully understood, we allowed a free relative normalization of the HEXTE data with respect to the PCA data. In all our fits we discard the PCA data below 4 keV.

The {\it CGRO}/OSSE observation from June 14--25 (from 50 keV to 1 MeV) overlaps with the \xte\ observations on June 17 and 18. We have extracted six OSSE data sets near simultaneous with the \xte\ observations. In order to get better statistics we have increased each OSSE data interval to include 30 minutes on either side of the corresponding \xte\ interval.

\asca\/ observed Cyg X-1 from May 30 5:30 (UT) through May 31 3:20 (Dotani at al.\ 1997). The GIS observation was made in the standard PH mode. The SIS data suffer from heavy photon pile-up and thus are not usable. We have selected 1488 live seconds of the \asca\/ data near simultaneous with the corresponding \xte\ observation.

A simple, phenomenological model fitted to the \xte/OSSE observations of June 17--18 consists of a blackbody and a power-law continuum with an energy index, $\alpha$, around 1.3--1.5 and a Compton reflection component (Gierli\'nski et al. 1997). Since there is no apparent high-energy cutoff in the power law, we assume the non-thermal origin of the X/\g\ continuum. Hence, in this paper we fit the data with the hybrid thermal/non-thermal model (Coppi, Zdziarski and Madejski 1998). The numerical code embodies Compton scattering, two-photon pair production, pair annihilation, cooling of pairs via Coulomb scattering, e--e bremsstrahlung, and synchrotron radiation. All distributions are assumed to be isotropic and homogeneous. The seed photons have a blackbody spectrum of a temperature $T_{\rm s}$, and enter the hot plasma at a rate characterized by the soft photon compactness \ls $= {\sigma_{\rm T} L_{\rm s}}/{m_{\rm e}c^3R}$. The electrons and pairs are powered from external sources both by thermal heating and non-thermal acceleration. The non-thermal source has a power-law distribution $N(\gamma) \propto \gamma^{-\Gamma_{\rm inj}}$. The thermal compactness, \lth, and the non-thermal one, \lnth, measure the heating and acceleration rates, respectively. We find that compactness ratios are more convenient to use than compactness parameters themselves. Thus we employ \ls, \lh/\ls\ and \lnth/\lh\ in our fits, where \lh\ $\equiv$ \lth\ + \lnth.

Another parameter is the optical depth to scattering on ionization electrons, \taup. The total optical depth (including scattering on pairs), \tauT, and the equilibrium temperature of the thermal component, $T_{\rm th}$, are computed self-consistently. A fraction of $\Omega/2\pi$ of photons emerging from the hot source gets Compton-reflected from a cold, ionized accretion disk. To model reflection we use the angular-dependent Green's functions (Magdziarz and Zdziarski 1995). We also include the iron fluorescent line in our model. When the reflection comes from a fast rotating disk in the vicinity of a black hole, Doppler and gravitational shifts become important. We approximate these effects convoluting the reflected component with the disk line profile in the Schwarzschild metric (Fabian et al. 1989). We assume the cold disk extends from $R_{\rm in} = 3R_{\rm g}$ to $R_{\rm out} = 500R_{\rm g}$ (where $R_{\rm g} = 2GM/c^2$), and is inclined to the observer at angle 36$^\circ$ (Davis and Hartmann 1983).

\section{RESULTS}

\subsection{Hot plasma}

The data can hardly constrain absolute values of compactness parameters. However, from the fits we can limit the ratio of \lh/\ls. This and the total X/\g\ luminosity (assuming distance to the source of 2.5 kpc) lead to an upper limit on the soft photon compactness of $l_{\rm s,max} \sim (10^9$~cm)/$R$, where $R$ is the size of the hot plasma. Since we expect a corona covering an internal part of the cold disk up to at least $\sim$ 10\Rg\ (see section 2.3), we can constrain \ls$\simless$30. The high \ls\ is also unlikely due to lack of an annihilation line (see below in this section). We have performed several fits with \ls\ fixed at various values and found a weak dependence of \chisq\ on \ls\ for high compactness, but decreasing \ls\ below $\sim$ 5 results in an abrupt rise of \chisq. Hence, in all our fits we fix \ls = 5.

\begin{figure}[t]
\vspace{9pt}
\epsfxsize=7.4cm \epsfbox{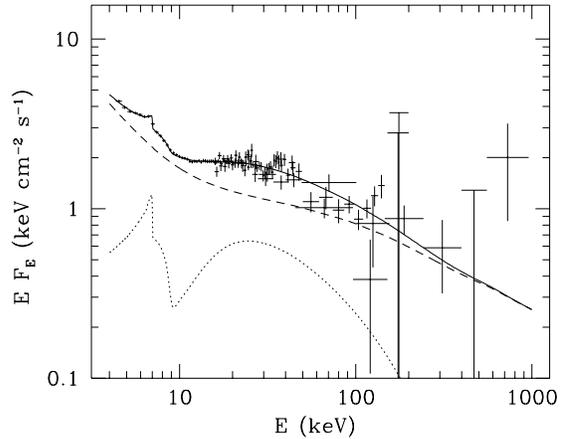}
\caption{The \xte/OSSE observation of Cyg X-1 in the soft state, on June 17, 1996. The dashed curve shows the continuum, the dotted one -- the component reflected from the cold disk. Solid curve shows the sum. The best fit parameters are: \lh/\ls\ = 0.34, \lnth/\lh\ = 0.92, \taup\ = 0.4, \kTth\ = 38 keV, \O2p\ = 0.8, $\chi^2$ = 424 / 474 d.o.f.}
\label{fig:juneobs}
\end{figure}

First we fit simultaneous \xte/OSSE data of June 17--18. Since we do not have reliable data below 4 keV, we cannot fit the seed photons temperature, \Ts, so we keep it at \kTs\ = 300 eV. We take advantage of the broad energy range (4--1000 keV) to constrain plasma properties. A pure thermal plasma (as fitted to the soft state data by Cui et al.\ 1997) is not compatible with the data, yielding an unacceptable \chisq\ of $\sim$1000 at 475 degrees of freedom. This is because thermal Comptonization is not capable of producing a power-law spectrum without a high-energy cutoff. Therefore, the significant contribution from non-thermal electrons (pairs) is essential. The best-fit value of \lnth/\lh\ is 0.9--1, which implies that thermal heating is weak. We note that the case of \lnth/\lh\ = 1 does not mean the Comptonizing electrons (pairs) are purely non-thermal. The non-thermal electrons (pairs) can transfer significant fraction of their energy to the thermal ones via Coulomb cooling. The best fit value to the optical depth, \taup, is about 0.4, and we find \kTth\ $\sim$ 40 keV, which corresponds to the Compton $y$ parameter of about 0.13. The optical depth to scattering on non-thermal electrons (pairs) is of order of $10^{-2}$. The thermal electrons (pairs) contribute mainly to the low-energy part of the emerging X-ray spectrum, while the high-energy tail is produced by a single scattering on power-law electrons (pairs). Figure 1 shows the model fitted to the first of the six June 17--18 observations.

Subsequently, we study the \asca/\xte\ data of May 30. With the seed blackbody photons with \kTs\ = $321^{+4}_{-5}$ eV (see section 2.3) we get similar plasma parameters as for June 17--18 data. We found \lh/\ls\ = $0.54^{+0.02}_{-0.03}$, \lnth/\lh\ = $0.94^{+0.06}_{-0.03}$, \taup\ = $0.34^{+0.03}_{-0.01}$ and \Gin\ = $2.68^{+0.09}_{-0.03}$. These parameters imply \tauT\ = 0.35 and \kTth\ = 49 keV.

\begin{figure}[t]
\vspace{6pt}
\epsfxsize=7.4cm \epsfbox{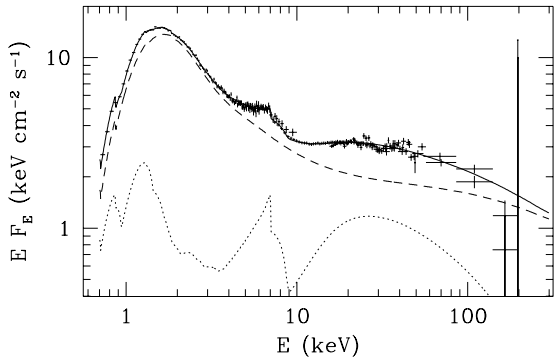}
\caption{The \asca/\xte\ observation of Cyg X-1 in the soft state, on May 30, 1996. The dashed curve shows the continuum, the dotted one - the component reflected from the cold disk. Solid curve shows the sum. The model details and best fit values are described in the text.}
\label{fig:mayobs}
\end{figure}

The next issue is presence of electron-positron pairs in the source. In all our fits \taup $\approx$ \tauT, i.e.\ the contribution from pairs to the optical depth is negligible, and plasma is proton-dominated. We have investigated the case of pure pair plasma (\taup = 0), but the obtained \chisq\ is much worse (by 100--200) for the assumed \ls\ = 5. The acceptable fit can be achieved only for high compactness values, e.g.\ for \ls\ = 50. In that case the data do not allow distinguishing univocally between pair and electron plasma. Nonetheless, the high compactness implies the presence of an annihilation line not observed in the high state of Cyg X-1, neither in the averaged OSSE data of June 14--25 1996, nor in the OSSE observations of a similar state in February 1--8, 1994 (Phlips et al. 1996). The high compactness also leads to extremely small size of the corona. Hence, we find the presence of pairs in the soft state of Cyg X-1 unlikely.

\subsection{Compton reflection}

We find the presence of Compton reflection from the cold matter in the soft state of Cyg X-1. We have examined the importance of Doppler and gravity shifts, and found that model with the relativistic reflection yields significantly better fit than that the static one ($\Delta \chi^2 \sim 18$/572 d.o.f.). The reflector covers the solid angle of $\Omega \sim 0.8\times 2\pi$ for the June 17--18 observations and $0.71^{+0.06}_{-0.09}\times 2\pi$ for May 30 data. The reflecting medium is highly ionized, with $\xi = (1.36^{+0.12}_{-0.23})\times10^4$ erg cm s$^{-1}$ (\asca/\xte\ data of May 30), which corresponds to hydrogen-like Fe {\sc xxvi} as a most abundant ($\sim$ 50\%) iron species. The result is comparable to that obtained by Cui et al. (1997) but substantially higher then $\xi = 430^{+160}_{-130}$ erg cm s$^{-1}$ found by Gierli\'nski et al. (1997), where the authors modeled continuum as a power law. The discrepancy shows that modeling Comptonization by a simple power law may lead to significant inaccuracies.

From the derived ionization state we would expect fluorescence line from Fe~{\sc xxvi} at $\sim$ 7 keV and some contribution from Fe~{\sc xxv} at $\sim$ 6.7 keV. The best-fit value of the rest-frame line energy is $6.59^{+0.16}_{-0.18}$ keV for the assumed inclination angle of 36$^\circ$, but may be as high as $6.77^{+0.23}_{-0.18}$ for $i$ = 30$^\circ$. We note that estimating the rest-frame line energy is dependent on various geometrical presumptions, namely on the postulated metric (Schwarzschild or Kerr), disk inclination and a line emissivity along the disk radius.

\subsection{Cold accretion disk}

The X-ray spectrum of Cyg X-1 in the soft state below $\sim$ 3 keV is dominated by the soft component attributed to the emission from the cold accretion disk. The disk spectrum is usually approximated by a multi-color disk model (MCD, Makishima et al. 1986), characterized by the inner disk radius, \Rin, and the temperature $T_{\rm in} = T(R_{\rm in})$. We model the disk spectrum by a single-temperature blackbody, at the same time taking it as a source of the seed photons for Comptonization. For the photon energy $E \simgreat 2kT_{\rm in}$ the MCD spectrum is well approximated by a single blackbody with a temperature $T_{\rm s} = 0.7T_{\rm in}$. The best-fit value of \kTs\ for the May 30 observation is $321^{+4}_{-5}$ eV, which is consistent with the inner disk temperature, \kTin\ = 430 eV, found by Dotani et al.\ (1997) and Cui at al.\ (1997). The spectrum is absorbed by a hydrogen column of $N_{\rm H}$ = $0.38^{+0.02}_{-0.01}$.

An important issue is the stability of such a disk. Let us define the relative accretion rate as $\dot{m} = \eta \dot{M}c^2 / L_{\rm Edd}$, where $\eta = 0.06$ is accretion efficiency in the Schwarzschild metric, and \Ledd\ is Eddington luminosity. The standard Shakura-Sunayev model implies a critical accretion rate $\dot{m}_{\rm crit} = 0.022(\alpha m)^{-1/8}(1-f)^{-9/8}$, where $m = M/M_{\odot}$ and $f$ is a fraction of accretion energy transferred from the disk to a corona (Svensson and Zdziarski 1995). For $\dot{m} > \dot{m}_{\rm crit}$ a radiation-pressure dominated region exists in the disk, which is thought to cause an instability. For $m \approx 10$ and assumed $\alpha = 0.1$, $\dot{m}_{\rm crit} \approx 0.02(1-f)^{-9/8}$. The observed $\dot{m} \approx$ 0.04 is below the critical accretion rate only when non-negligible fraction of accretion power ($f \simgreat 0.4$) is dissipated in the corona, covering the inner unstable region of the cold disk up to at least $\sim$ 10\Rg. The presence of the corona is consistent with observations, where the cold disk probably extends to the last stable orbit, the high-energy spectrum is produced by the optically thin plasma, and cold reflector covers substantial angle of $\sim 0.7 \times 2\pi$. From the energy balance we estimate $f$ to be around 0.4, which places Cyg X-1 at the end of the stable solution branch. Figure 3 shows the solutions of the optically thick accretions disk with the corona for $f$ = 0, 0.4 and 0.9. 

\begin{figure}[t]
\epsfxsize=7.4cm \epsfbox{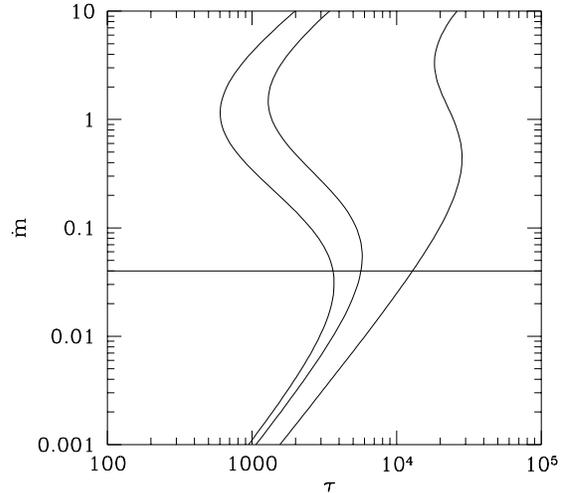}
\caption{The solutions of the optically thick disk with a fraction of accretion power, $f$, dissipated in the corona, for $m$ = 10 and $\alpha$ = 0.1, at a radius of the most significant instability $R = 5.72R_{\rm g}$ (Svensson and Zdziarski 1995). Three curves from the left to the right correspond to $f$ = 0, 0.4 and 0.9. The horizontal line corresponds to the accretion rate of Cyg X-1 in the soft state.}
\label{fig:mdottau}
\end{figure}

\end{document}